\documentclass{article}
\usepackage{spconf}
\ninept
\usepackage{tikz}
\usepackage{amsmath,amsfonts,amssymb}%
\usepackage{bm}%
\usepackage{graphicx,graphics}%
\usepackage{cases}%
\usepackage[noadjust]{cite}%
\usepackage{color}%

\usepackage{verbatim}
\usepackage{algorithm}
\usepackage{balance}
\usepackage{multirow}
\usepackage{stfloats}
\usepackage{xspace}
\usepackage{amsthm}
\usepackage{mathtools} 
\usepackage{subfig} 

\usepackage{caption}

\usepackage{tikz}
\usetikzlibrary{shapes.misc}
\usetikzlibrary{matrix}
\usetikzlibrary{arrows,backgrounds,fit,calc}

\newcommand{\secref}[1]{Section\,\ref{#1}}

\DeclarePairedDelimiterX\abs[1]{\lvert}{\rvert}{#1}
\DeclarePairedDelimiterX\parn[1]{(}{)}{#1}
\DeclarePairedDelimiterX\set[1]{\lbrace}{\rbrace}{#1}
\DeclarePairedDelimiterX\innerp[2]{\langle}{\rangle}{#1,#2}
\DeclarePairedDelimiterX\norm[1]{\lVert}{\rVert}{#1}
\DeclarePairedDelimiterX\brak[1]{\lbrace}{\rbrace}{#1}
\DeclarePairedDelimiterX\coeff[1]{(}{)}{#1}

\newcommand{\conj}[1]{\overline{#1}} 
\newcommand{\untsph}{\mathbb{S}^{2}} 

\newcommand{\lsph}{L^{2}(\mathbb{S}^{2})}

\newcommand{\nps}{\ensuremath{N_\textrm{O}}}

\newcommand{\bv}[1]{\boldsymbol{#1}}

\newcommand{\intsph}{\int_{\mathbb{S}^{2}}}
\newcommand{\intphi}{\int_{0}^{2\pi}}
\newcommand{\plms}{\widetilde{P}}
\newcommand{\dfn}{\triangleq}
\newcommand{\figref}[1]{Fig.\,\ref{#1}}

\newcommand{\shc}[3]{\coeff{#1}{}_{#2}^{#3}}

\newtheorem{remark}{Remark}

\graphicspath{{figs/},{Figures/}}

\newcommand{\displayskipshrink}{%
    \setlength{\abovedisplayskip}{3.0pt plus 1.0pt minus 1.0pt}
    \setlength{\abovedisplayshortskip}{0pt plus 1.0pt minus 1.0pt}
    \setlength{\belowdisplayskip}{3.0pt plus 1.0pt minus 1.0pt}
    \setlength{\belowdisplayshortskip}{3.0pt plus 1.0pt minus 1.0pt}}

\title{An Optimal Dimensionality Sampling Scheme on the Sphere for Antipodal Signals in diffusion magnetic resonance imaging}

\name{Alice P. Bates, Zubair Khalid and  Rodney A. Kennedy
\thanks{This work is supported by the Australian Research Council's Discovery Projects funding scheme (Project no. DP150101011).}}
\address{Research School of Engineering, The Australian National University, Canberra, ACT 2601, Australia\\
Email: \{alice.bates, zubair.khalid rodney.kennedy\}@anu.edu.au}

\begin{document}
\displayskipshrink

\maketitle

\begin{abstract}

We propose a sampling scheme on the sphere and develop a corresponding spherical harmonic transform~(SHT) for the accurate reconstruction of the diffusion signal in diffusion magnetic resonance imaging~(dMRI). By exploiting the antipodal symmetry, we design a sampling scheme that requires the optimal number of samples on the sphere, equal to the degrees of freedom required to represent the antipodally symmetric band-limited diffusion signal in the spectral~(spherical harmonic) domain. Compared with existing sampling schemes on the sphere that allow for the accurate reconstruction of the diffusion signal, the proposed sampling scheme reduces the number of samples required by a factor of two or more. We analyse the numerical accuracy of the proposed SHT and show through experiments that the proposed sampling allows for the accurate and rotationally invariant computation of the SHT to near machine precision accuracy.
\end{abstract}

\vspace{-1mm}
\section{Introduction}
\vspace{-1mm}
Diffusion magnetic resonance imaging (dMRI) allows for the structure and connectivity of white matter in the brain to be determined non-invasively for the detection of neurological diseases and in preoperative planning \cite{Basser:1994,Jahanian:2008}. White matter has a fibrous structure; the myelin sheath around fibre axons hinders the movement of water molecules,
constraining diffusion. The diffusion of water molecules in white matter can therefore be used to determine the location and orientation of fibres.



Acquisition and reconstruction of the diffusion signal from $\bv q$-space measurements, where $\bv q$ is the diffusion wave vector, is a central problem in dMRI. It is common for samples to be obtained on a single sphere or multiple spheres in $\bv q$-space~\cite{cheng:2014, ye:2012, daducci:2011}. The number of measurements in $\bv q$-space~(images) that can be acquired in dMRI is severely limited due to the need for the scan time to be practical in a clinical setting. Hence, a $\bv q$-space sampling scheme that allows for the accurate reconstruction of the diffusion signal using the minimum number of samples is desirable \cite{assemlal:2009,tournier:2013}.


The reconstruction of the diffusion signal can be carried out by expanding the signal in terms of spherical harmonics~\cite{haldar:2012,descoteaux:2007,anderson:2005}, which serve as natural basis functions for signals defined on the sphere~\cite{Kennedy-book:2013}.
By choosing a sufficiently large band-limit, the diffusion signal can be represented in terms of a finite number of spherical harmonic coefficients, which are obtained using the spherical harmonic transform~(SHT) calculated numerically from a finite number of measurements of the diffusion signal obtained at a constant $\bv q$-space radius (on a sphere)~\cite{tournier:2013, hess:2006}. In order for the accurate reconstruction of the diffusion signal, a sampling scheme on the sphere for obtaining $\bv q$-space measurements needs to allow for the accurate computation of the SHT.


\vspace{-1mm}
\subsection{Relation to Prior Work}
Many sampling schemes on the sphere in dMRI use an antipodally symmetric sampling grid to reduce the number of measurements; the antipodal symmetry of the diffusion signal
enables the signal to be evaluated on the sphere at locations antipodal to where measurements have been obtained~\cite{jones:1999,descoteaux:2007,tuch:2004,fejes-toth:1949,papadakis:2000,caruyer:2013}. Most of these schemes are designed to obtain uniform~(near equal area) sampling on the sphere to ensure that the reconstruction is rotationally invariant, which means that the accuracy does not vary significantly if the sampling grid (or diffusion signal) is rotated \cite{cheng:2014,caruyer:2013}. Other sampling scheme designs that focus on the accurate computation of the SHT rather than geometric properties of the diffusion signal have also been used~\cite{daducci:2011}.


In order to acquire samples of the diffusion signal band-limited at $L$~(formally defined in \secref{subsec1:math_prelims}) over multiple spheres in $\bv q$-space, an equiangular scheme~\cite{McEwen2011} that requires asymptotically $2L^2$ samples to accurately compute the SHT has been used~\cite{daducci:2011}. In comparison to the geometric schemes, this scheme~\cite{McEwen2011} has an iso-latitude arrangement of samples~(where samples along longitude are taken over iso-latitude annuli) that enables the fast computation of the SHT. However, the samples in the scheme are non-uniform due to dense sampling near the poles, therefore it is considered that the accuracy of the SHT changes if the signal is rotated~\cite{caruyer:2013,cheng:2014}. The spherical design with uniform density sampling method in \cite{caruyer:2013} has a uniform and antipodally symmetric arrangement of samples, and allows for the accurate computation of the SHT using least squares. The number of samples required by this scheme does not have a general formula depending on the band-limit but as reported in ~\cite{caruyer:thesis} is greater than $L^2$.

For the accurate computation of the SHT, the minimum number of samples, referred as the optimal spatial dimensionality, required by any sampling scheme is equal to the number of spherical harmonic coefficients required to represent a band-limited signal in spherical harmonics basis~\cite{McEwen2011,Khalid:2014}. An iso-latitude scheme that attains optimal spatial dimensionality ($L^2$ samples) of an arbitrary band-limited signal on the sphere has been developed~\cite{Khalid:2014}, which allows the accurate and fast computation of the SHT. However, the samples in the sampling scheme are neither antipodally symmetric nor uniform~(equal areal) by design. Due to antipodal symmetry of the diffusion signal, the band-limited diffusion signal has $\nps = L(L+1)/2$ degrees of freedom~(see \secref{Sec:prob_formulation}), which is referred to as the optimal spatial dimensionality of the diffusion~(antipodal) signal. We have established that the existing sampling schemes require at least $L^2 \approx 2\nps$ samples on the sphere for the accurate computation of the SHT~(or the accurate reconstruction) of the diffusion signal.

\vspace{-1mm}
\subsection{Contributions}

In this work, we address whether it is possible to develop a sampling scheme on the sphere (at a constant $\bv q$-space radius) for measuring the diffusion signal that 1) attains the optimal spatial dimensionality, i.e., requires $\nps$ samples, 2) allows for the accurate computation of the SHT up to and beyond practically relevant band-limits, 3) permits rotationally invariant reconstruction, and 4) enables fast computation of the SHT with lowest known complexity.
%
%
%
\noindent In addressing these questions, we develop a sampling scheme on the sphere and corresponding SHT that, in contrast to previous schemes, attains the optimal spatial dimensionality for the accurate reconstruction of the band-limited diffusion signal at a constant $\bv q$-space radius (an antipodally symmetric signal). The proposed scheme is antipodal, but unlike existing schemes that only focus on the geometric properties of the diffusion signal, the SHT associated with the proposed scheme is computationally efficient and accurate, with reconstruction error near machine precision ($10^{-14}$) for the band-limits of interest in dMRI. Though the scheme is non-uniform by design, we show that the numerical accuracy of the scheme is the same when the diffusion signal on the sphere is rotated, demonstrating that the scheme allows for the rotationally invariant reconstruction of the diffusion signal.

This paper is organised as follows. The mathematical preliminaries necessary to understand this work are contained in \secref{Sec:math_prelims}. In \secref{Sec:prob_formulation}, we state the problem under consideration. The sampling scheme and the SHT for measurement and reconstruction of the diffusion signal are presented in \secref{Sec:sampling_scheme}. This scheme is then evaluated in terms of numerical accuracy and rotational invariance in \secref{Sec:analysis}. Finally, concluding remarks are made in \secref{Sec:conclusions}.

\vspace{-1mm}
\section{Mathematical Preliminaries}\label{Sec:math_prelims}
\vspace{-1mm}


\subsection{Signals on the Sphere and Spherical Harmonics}
\label{subsec1:math_prelims}
Square integrable complex functions on the unit sphere $\untsph$ have the form $f(\theta,\phi)$, where the angles co-latitude $\theta \in [0, \pi]$ and longitude $\phi\in [0, 2\pi)$ parameterise a point $(\sin\theta\cos\phi,\, \sin\theta\sin\phi,\, \cos \theta)' \in \mathbb{R}^3$ on $\untsph$. The inner product of two functions $f$, $g$ defined on $\untsph$ is given by~\cite{Kennedy-book:2013}
\begin{align}\label{eqn:innprd}
\langle f, g \rangle \triangleq  \intsph
f(\theta,\phi) \overline {g(\theta,\phi)}
\,\sin\theta\, d\theta\, d\phi,
\end{align}
where $\overline{(\cdot)}$ denotes the complex conjugate and $\sin\theta\,d\theta\,d\phi$ is the differential area element on the sphere. With the inner product in (\ref{eqn:innprd}), the set of square integrable complex valued functions on $\mathbb{S}^2$ forms a Hilbert space, denoted by $\lsph$. The inner product in (\ref{eqn:innprd}) induces a norm $\|f\| \triangleq\langle f,f \rangle^{1/2}$. We refer to functions with finite induced norm as signals on the sphere.

The spherical harmonic functions (spherical harmonics for short) $Y_{\ell}^m(\theta, \phi)$, defined for integer degree ${\ell} \geq 0$ and integer order $ |m| \leq {\ell}$, form a complete orthonormal set of basis functions~\cite{Kennedy-book:2013, Sakurai:1994}.
The signal is said to be band-limited at degree $L$ if the signal can be \emph{completely} represented in spherical harmonics basis functions $Y_{\ell}^m(\theta,\phi)$ with $\ell<L$ and $|m|\le \ell$.

\subsection{Diffusion Signal on the Sphere}

The diffusion signal at a fixed $\bv q$-space radius can be represented as a band-limited signal on the sphere~\cite{caruyer:2013,hess:2006}. Furthermore, the diffusion signal, denote by $d(\theta,\phi)$, is antipodally symmetric, with $d(\theta,\phi) = d(\pi - \theta,\phi + \pi)$. Since $Y_{\ell}^m(\theta, \phi)=Y_{\ell}^m(\pi-\theta, \pi+\phi)$ for even $\ell$ and $Y_{\ell}^m(\theta, \phi)=-Y_{\ell}^m(\pi-\theta, \pi+\phi)$ for odd $\ell$, the expansion of $d(\theta,\phi)$ in spherical harmonic basis only includes even degree spherical harmonics, that is,

%
\begin{equation}
\label{Eq:f_expansion}
    d(\theta,\phi)=\sum_{\substack{\ell=0 \\ \ell \, \rm{even}}}^{L-1}\sum_{m=-{\ell}}^{\ell} (d)_{\ell}^{m}  Y_{\ell}^m(\theta,\phi), \quad L~{\rm odd},
\end{equation}
where $(d)_{\ell}^{m}$ denotes the spherical harmonic coefficient of degree $\ell$ and order $m$, and is calculated using the SHT, given by
\begin{equation}\label{Eq:fcoeff}
(d)_{\ell}^{m} \dfn \intsph d(\theta,\phi) \conj {Y_{\ell}^m(\theta,\phi)}
\,\sin\theta\, d\theta\, d\phi.
\end{equation}
The spherical harmonic coefficients $\shc{d}{\ell}{m}$ form the spectral domain representation of $d(\theta,\phi)$.
The reconstruction of the signal $d(\theta,\phi)$ from its spherical harmonic coefficients as given in \eqref{Eq:f_expansion} is referred as the inverse SHT.
The band-limit required to accurately represent the diffusion signal depends on the $\bv q$-space radius~\cite{daducci:2011}, and band-limits up to $L=11$ are typically used \cite{hess:2006,tournier:2013}.

\vspace{-1mm}
\section{Problem Formulation}\label{Sec:prob_formulation}
\vspace{-1mm}
The diffusion signal $d(\theta,\phi)$ has only even order spherical harmonic coefficients due to its antipodal symmetry. The number of spherical harmonics required to represent $d(\theta,\phi)$ in \eqref{Eq:f_expansion} is $\nps = L(L+1)/2$~\cite{tournier:2013,caruyer:2012}, which also represents the optimal dimensionality attainable by any sampling scheme that allows the accurate computation of the SHT of the diffusion signal.
We customise the recently developed sampling scheme \cite{Khalid:2014}, which requires the optimal number of samples for the accurate computation of the SHT of an arbitrary band-limited signal (without antipodal symmetry), for the acquisition and reconstruction of the diffusion signal in dMRI.


\subsection{Optimal Dimensionality Sampling Scheme}\label{Sec:struc_opt_dim}

The optimal dimensionality sampling scheme in \cite{Khalid:2014} has an iso-latitude sampling grid, with samples taken over $L$ iso-latitude rings. The $L$ locations along $\theta$ where the iso-latitude rings are placed are stored in the vector $\bv\theta$ defined as
\begin{align}\label{Eq:theta_general}
\bv\theta \dfn [\theta_0,\, \theta_1,\,\hdots,\,\theta_{L-1}]^T.
\end{align}
For the accurate computation of the SHT, there needs to be \emph{at least} $2n+1$ samples along longitude in the ring placed at $\theta_n \in \bv\theta$.
The iso-latitude rings are placed along $\theta$ such that the SHT can be computed accurately~(see \cite{Khalid:2014} for further details). For the accurate computation of the SHT, the number of samples required by the optimal spatial dimensionality scheme is $L^2$, which is equal to the number of degrees of freedom required to represent an arbitrary band-limited signal defined on the sphere.

\subsection{Research Questions}

In this work, we address the following questions:
\vspace{-0.5mm}
\begin{itemize}\itemsep0mm
\item How can we customise the structure of the optimal dimensionality sampling scheme to make it suitable for the acquisition of measurements of the antipodal diffusion signal?
\item Can we exploit the antipodal symmetry property of the diffusion signal to develop a scheme that requires only $\nps = L(L+1)/2$ rather than $L^2$ samples while still allowing for the accurate computation of the SHT ?
\end{itemize}

\vspace{-1mm}
\section{Proposed Sampling Scheme}\label{Sec:sampling_scheme}

\vspace{-1mm}
\subsection{Proposed Sampling Scheme --- Structure and Design}
The diffusion signal on the sphere is antipodally symmetric, with $d(\theta,\phi) = d(\pi - \theta,\phi + \pi)$. Hence, by measuring the signal at a location $(\theta_{A},\phi_{A}) $, we also know the value of the signal at a second location $(\theta_{B},\phi_{B}) = (\pi - \theta_{A},\phi_{A} + \pi)$. We use this property to design the sampling scheme in order to reduce the number of measurements of the diffusion signal that need to be taken.

With this consideration, we propose to place the iso-latitude rings in pairs such that they are antipodal to one another. The vector $\bv\theta$, given in \eqref{Eq:theta_general}, describing the location of the iso-latitude rings is
\begin{align}\label{Eq:theta_antipodal}
\bv\theta \dfn [0, \hdots,\, \pi - \theta_{L-3},\theta_{L-3},\pi - \theta_{L-1},\theta_{L-1}]^T,\quad L \,\,\textrm{ odd}.
\end{align}
 We shortly present the location of these samples along co-latitude. Since we require at least $2n+1$ samples along $\phi$ in the iso-latitude ring placed at $\theta_n$ for the accurate computation of the SHT, we propose equiangular sampling along longitude, with $k$-th sample location, denoted by $\phi^n_k$, in the ring placed at $\theta_n$ is given by
%
\begin{align}\label{Eq:phi_equally_spaced}
\phi^n_k \dfn \begin{cases}\frac{2k\pi}{2n+1},& n =0,\,2,\,\hdots,\,L-1, \quad k \in[0,\, 2n], \\
\frac{\pi(2k+1)}{2n+3},& n =1,\,3,\,\hdots,\,L-2, \quad k \in [0,\,2(n+1)].
\end{cases}
\end{align}

\begin{figure}[t]
  \centering
  \vspace{-2mm}
  \subfloat[]{
  \includegraphics[width=0.17\textwidth]{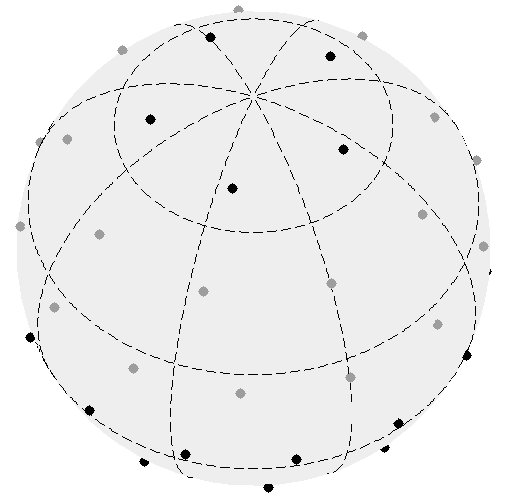}}
  \hspace{10mm}
   \subfloat[]{
  \includegraphics[width=0.17\textwidth]{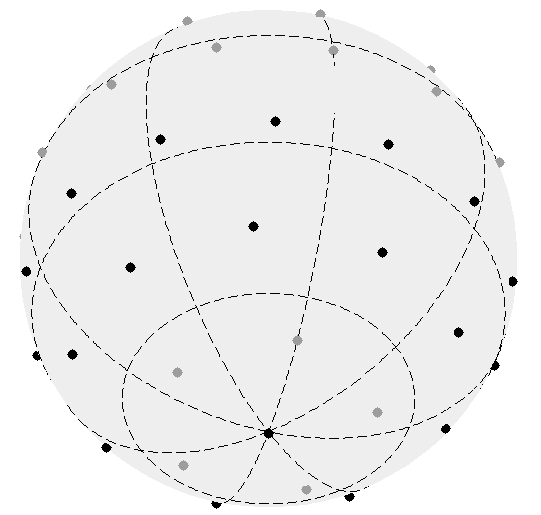}}
  \vspace{-2mm}
  \caption{(a) North pole view and (b) South pole view of the proposed sampling scheme on the sphere given by \eqref{Eq:theta_antipodal} and \eqref{Eq:phi_equally_spaced} for measuring $d(\theta,\phi)$ band-limited at $L = 7$. The locations where measurements need to be taken are shown in black and the locations where $d(\theta,\phi)$ is evaluated using antipodal symmetry are shown in grey.}
  \label{fig:sampling_grid}
  \vspace{-2mm}
\end{figure}

\begin{remark}
The samples in the proposed scheme are structured in a way that the samples in the ring $\theta_n$ are antipodal to the samples in the ring $\theta_{n-1}$ for $n =2,\,4,\,\hdots,\,L-1$. Therefore, we only need to take the measurements over the rings $\theta_n$ for $n =0,\,2,\,\hdots,\,L-1$. The values of the signal over the remaining samples can be determined by using the antipodal symmetry of the diffusion signal and exploiting the structure of the proposed sampling scheme. As an example, \figref{fig:sampling_grid} shows the proposed sampling scheme for $L=7$.
\end{remark}
%


\begin{remark}
Since the measurements are only required to be taken over $(L+1)/2$ rings in the proposed scheme, the total number of measurements required by the proposed sampling scheme is
\begin{equation}\label{Eq:number_samples_antipodal}
\sum_{\substack{n=0 \\ n {\rm even}}}^{L-1}(2n+1) = \frac{(L+1)L}{2} = \nps,
\end{equation}
which also represent the degrees of freedom required to represent band-limited antipodal signal. Thus, the proposed scheme attains the
optimal spatial dimensionality.
\end{remark}

\vspace{-1mm}
\subsection{Spherical Harmonic Transform (SHT) --- Formulation} \label{sec:SHT}
We here present a variant of the method developed in~\cite{Khalid:2014}, to compute the SHT of the diffusion signal sampled using the proposed sampling scheme. We define an indexed vector $\bv\theta^{m} \dfn [\theta_{|m|},\, \theta_{|m|+1},\,\hdots,\,\theta_{L-1}]^T \subset\bv{\theta}$ for order $|m| < L$, which contains the last $L-|m|$ points in the vector $\bv{\theta}$.
%
%
%
Also define a vector
\begin{equation}
\nonumber
\mathbf{g}_m \equiv G_m(\bv\theta^{m}) \dfn [G_m(\theta_{|m|}),\,G_m(\theta_{{|m|}+1}),\,\hdots,\, G_m(\theta_{L-1})]^T,
\end{equation}
with
\begin{equation}\label{Eq:gm_integral}
G_m(\theta_n) \dfn \intphi \! f(\theta_n,\phi) e^{-im\phi} d\phi = 2\pi \sum_{\ell=|m|}^{L-1} \shc{f}{\ell}{m}  \plms_\ell^m(\theta_{n}),
\end{equation}
for each order $|m|<L$ and each $\theta_n\in \bv{\theta}$, where $\plms_\ell^m(\theta_{n}) \dfn Y_\ell^m(\theta_{n},0)$ denotes scaled associated Legendre functions.

If $G_m(\theta_n)$ for each $\theta_n\in\bv\theta^m$ is computed, the spherical harmonic coefficients of order $m$ can be recovered from \eqref{Eq:gm_integral} by setting up a system of linear equations~\cite{Khalid:2014}, given by,
\begin{equation}\label{Eq:gtof}
\mathbf{g}_m  = \mathbf{P}_L^m \mathbf{f}_m,\, \quad |m|\le L,
\end{equation}
where $\mathbf{P}_L^m$ is a matrix of size $(L-m)\times(L-m)$ defined as
\begin{equation*}
\mathbf{P}_L^m \dfn 2\pi \begin{small}\setlength{\arraycolsep}{1mm}
\begin{pmatrix}
   \plms_{|m|}^m(\theta_{|m|}) & \plms_{{|m|}+1}^m(\theta_{|m|}) & \cdots & \plms_{L-1}^m(\theta_{|m|}) \\[2mm]
   \plms_{|m|}^m(\theta_{{|m|}+1}) & \plms_{{|m|}+1}^m(\theta_{{|m|}+1}) & \cdots & \plms_{L-1}^m(\theta_{{|m|}+1}) \\
   \vdots  & \vdots  & \ddots & \vdots  \\
   \plms_{|m|}^m(\theta_{L-1}) & \plms_{{|m|}+1}^m(\theta_{L-1}) & \cdots & \plms_{L-1}^m(\theta_{L-1})\\
  \end{pmatrix}\end{small},
\end{equation*}
and $\mathbf{f}_m$ is a vector composed of spherical harmonic coefficients of order $|m| < L$, given by
\begin{equation}
\mathbf{f}_m  = \big[ \shc{f}{{|m|}}{m},\,  \shc{f}{{|m|}+1}{m},\,\hdots,\, \shc{f}{L-1}{m}\big]^T.
\end{equation}

\vspace{-2mm}
\subsection{Spherical Harmonic Transform --- Computation}
\label{sec:SHT_computation}
The spherical harmonic coefficients of order $|m|<L$ contained in a vector $\mathbf{f}_{m}$ can be computed accurately by inverting \eqref{Eq:gtof}, provided $\bv\theta^m$ is chosen such that $\mathbf{P}_L^m$ is well-conditioned and $G_m(\theta_n)$ for each $\theta_n\in\bv\theta^m$ is computed accurately. Since we have considered in the design of the proposed scheme that the ring placed at $\theta_n$ contains at least $2n+1$ samples points along longitude, $G_m(\theta_n)$ for each $\theta_n\in\bv\theta^m$ can be \emph{exactly} computed by evaluating the integral equation in \eqref{Eq:gm_integral} as a summation over $2n+1$ samples along longitude~\cite{Khalid:2014}.


We also require the sampling points along co-latitude to be chosen such that the matrix $\mathbf{P}_L^m$ is well-conditioned for each $|m|\le L$. Since the locations of the iso-latitude rings along co-latitude in the vector $\bv\theta$ are required to be in pairs, as given in \eqref{Eq:theta_antipodal}, we are only required to obtain measurements of the diffusion signal at $(L+1)/2$ positions~($\theta_0,\,\theta_2,\,\hdots,\,\theta_{L-1}$) along co-latitude. We define a set of equiangular $(L+1)/2$ samples along co-latitude given by
\begin{equation}\label{Eq:equiangular_set}
\Theta = \bigg\{\frac{\pi(2t+1)}{2L-1} \bigg\}, \quad t=0,1,\hdots,\frac{L-1}{2},
\end{equation}
and propose the following method to determine the optimal ordering of samples in the vector $\bv{\theta}$:
\begin{itemize}
\item Choose $\theta_{L-1}=\frac{\pi(2\lfloor\frac{L-1}{2}\rfloor+1)}{2L-1}$ farthest from the poles, which is a natural choice for the ring of $2L-1$ (the largest number of) samples along $\phi$. Then, to satisfy the antipodal nature of the scheme, $\theta_{L-2}=\pi - \frac{\pi(2\lfloor\frac{L-1}{2}\rfloor+1)}{2L-1}$.
\item For each $m=L-3,\,L-5,\,\hdots 2$, choose $\theta_{m}$ and $\theta_{m-1} = \pi - \theta_{m}$ from the set $\Theta$, given in \eqref{Eq:equiangular_set}, that minimises the sum of the condition numbers of the matrices $\mathbf{P}^m$ and $\mathbf{P}^{m-1}$.
\item Choose $\theta_0 = 0$ or $\theta_0 = \pi$.
\end{itemize}
Such placement of samples along co-latitude ensures that $\mathbf{P}_L^m$ is well-conditioned, resulting in the accurate computation of the proposed SHT as we demonstrate in the next section.

We have considered the selection of sample positions along co-latitude from the set of equiangular samples as it has been shown in \cite{Khalid:2014} that the selection from the equiangular set results in more accurate computation of the SHT. Furthermore, the equiangular placement of samples along longitude, given in \eqref{Eq:phi_equally_spaced}, permits the use of the fast Fourier transform for the computation of $G_m(\theta_n)$ \eqref{Eq:gm_integral}.
We note that the proposed sampling scheme can be easily customised for the case when the samples along co-latitudes and/or longitude are not equiangular.
\vspace{-1mm}
\begin{remark}
The computational complexity of the SHT for the proposed sampling scheme is ($O(L^4)$)~\cite{Khalid:2014}, which is much smaller than the complexity of least squares methods ($O(L^6)$) used by many existing sampling schemes \cite{cheng:2014,caruyer:2013,koay:2011}.
\end{remark}
\vspace{-2mm}
\begin{figure}[!t]
    \centering
    \vspace{-2mm}
    \includegraphics[width=0.415\textwidth]{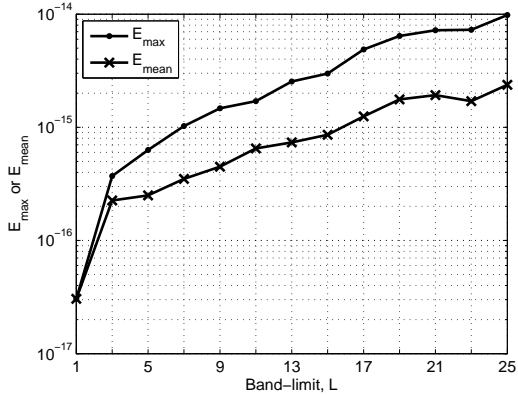}
    \vspace{-2mm}
    \caption{Plots of the maximum error $E_\textrm{max}$ and the mean error $E_\textrm{mean}$, given in \eqref{Eq:exp1:errors:max} and
    \eqref{Eq:exp1:errors:mean} respectively, for band-limits $L \leq 25$.}
    \vspace{-3mm}
    \label{fig:accuracy}
\end{figure}

\vspace{-2mm}
\section{Analysis of Proposed Sampling Scheme}\label{Sec:analysis}
\vspace{-2mm}
\subsection{Numerical Accuracy}\label{Sec:num_accuracy}

For the accurate reconstruction of the diffusion signal, using its expansion in the spherical harmonic basis in \eqref{Eq:f_expansion}, the accuracy of the SHT given in \eqref{Eq:fcoeff} is of the upmost importance. A sampling scheme is numerically accurate if the inverse SHT of a band-limited signal followed by the SHT gives an error between the original and reconstructed signal on the order of the numerical precision~\cite{Khalid:2014,McEwen2011}.

We carry out the following experiment to test the numerical accuracy of the proposed sampling scheme and the corresponding SHT. We first obtain a band-limited antipodally symmetric test signal $f_{\rm t}$ in the spectral domain by generating spherical harmonic coefficients $\shc{f_{\rm t}}{\ell}{m}$ for $0<\ell<L,  \: \ell  \, {\rm even}, \:  |m|\leq\ell$ with real and imaginary parts uniformly distributed in the interval $[-1,\,1]$ ($\shc{f_{\rm t}}{\ell}{m} = 0$ for $0<\ell<L,  \: \ell  \, {\rm odd}, \: |m|\leq\ell$). The inverse SHT is then used to obtain $f_{\rm t}$ in the spatial domain over the proposed sampling grid (described in \secref{Sec:sampling_scheme}), followed by the forward SHT to compute the spherical harmonic coefficients of the reconstructed signal, $\shc{f_{\rm r}}{\ell}{m}$. The experiment is repeated 10 times and the average values for the maximum error $E_\textrm{max}$ and the mean error $E_\textrm{mean}$, given by
\begin{align}\label{Eq:exp1:errors:max}
E_\textrm{max} &\dfn \max|\shc{f_{\rm t}}{\ell}{m} - \shc{f_{\rm r}}{\ell}{m} |, \\ E_\textrm{mean}  &\dfn \frac{1}{L^2} \sum_{\ell=0}^{L-1} \sum_{m=-\ell}^\ell |\shc{f_{\rm t}}{\ell}{m} - \shc{f_{\rm r}}{\ell}{m}| \label{Eq:exp1:errors:mean},
\end{align}
are recorded and plotted in \figref{fig:accuracy} for band-limits $ L\leq 25$ (the band-limits of interest in dMRI). It is evident that, although the errors $E_\textrm{max}$ and $E_\textrm{mean}$ grow as the band-limit $L$ increases, the errors are on the order of numerical precision (less than $10^{-14}$), showing that the proposed sampling scheme and SHT allows for the accurate reconstruction of any band-limited antipodally symmetric signal on the sphere for band-limits $L \leq 25$.

\vspace{-2mm}
\subsection{Rotational Invariance}
\begin{figure}[!t]
    \centering
    \vspace{-2mm}
    \includegraphics[width=0.42\textwidth]{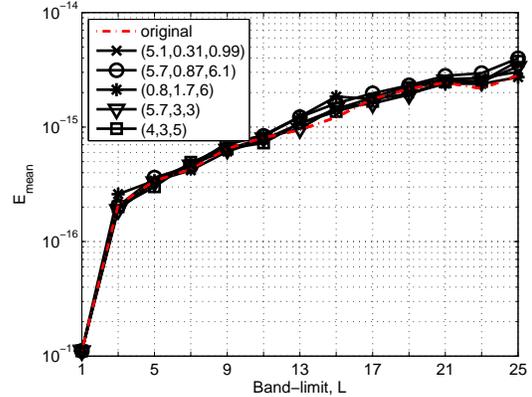}
    \vspace{-2mm}
    \caption{Plot of the mean error $E_\textrm{mean}$, given in \eqref{Eq:exp1:errors:mean}, for the original test signal $f_{\rm t}$ and 5 rotated versions of the test signal, and for band-limits $L \leq 25$. The legend displays the rotation angles in radians in the form of Euler angles $(\alpha,\beta,\gamma)$, where the rotation is applied to the test signal using the $zyz$ rotation convention~\cite{Kennedy-book:2013}. Note that the reconstruction accuracy is the same for all rotations.}
    \vspace{-3mm}
    \label{fig:rot_invariant}
\end{figure}

Many existing sampling schemes on the sphere focus on uniform sampling of the diffusion signal on the sphere to ensure that the reconstruction accuracy is rotational invariant i.e. does not change significantly when the signal or sampling grid is rotated \cite{caruyer:2013}. The proposed sampling scheme is not uniform by design, however it does not have dense sampling on any region of the sphere. The rotational invariance of the reconstruction accuracy of the SHT was tested using the numerical accuracy experiment (described in \secref{Sec:num_accuracy}) performed on 5 rotated versions of the antipodally symmetric test signal $f_{\rm t}$. For each rotation, we randomly obtain Euler angles $(\alpha,\beta,\gamma)$ from the uniform distributions, where $\alpha,\gamma \in [0,2\pi)$ and $\beta \in [0,\pi]$, and apply the rotation using the $zyz$ convention~\cite{Kennedy-book:2013}. \figref{fig:rot_invariant} shows the mean reconstruction error $E_\textrm{mean}$, averaged over 10 experiments, for the original test signal and 5 rotated versions of the test signal, for band-limits $L \leq 25$. It can be seen that the reconstruction errors are on the same order of magnitude regardless of the rotation angle, demonstrating that the accuracy of the scheme does not depend on the angle of rotation.

%
\vspace{-2mm}
\section{Conclusions} \label{Sec:conclusions}
\vspace{-1mm}
In this work we have proposed a new sampling scheme on the sphere for the accurate reconstruction of the diffusion signal in dMRI.
The proposed scheme exploits the antipodal symmetry of the diffusion signal and attains an optimal number of samples $\nps = (L+1)/2$, given by the degrees of freedom required to represent the diffusion signal of band-limit $L$, in contrast to the existing schemes which require at least $L^2$ samples. The smaller number of samples required by this scheme will enable a reduction in the scan time required to obtain sufficient measurements for the accurate reconstruction of the diffusion signal. The proposed sampling scheme enables accurate and rotational invariant computation of the SHT of the diffusion signal to near machine precision accuracy. Applying the scheme to dMRI data and extending the proposed scheme to multiple $\bv q$-shell sampling is the subject of our future work.

\bibliography{ICASSP_dMRI_Database} 

\begin{thebibliography}{10}
\providecommand{\url}[1]{#1}
\csname url@samestyle\endcsname
\providecommand{\newblock}{\relax}
\providecommand{\bibinfo}[2]{#2}
\providecommand{\BIBentrySTDinterwordspacing}{\spaceskip=0pt\relax}
\providecommand{\BIBentryALTinterwordstretchfactor}{4}
\providecommand{\BIBentryALTinterwordspacing}{\spaceskip=\fontdimen2\font plus
\BIBentryALTinterwordstretchfactor\fontdimen3\font minus
  \fontdimen4\font\relax}
\providecommand{\BIBforeignlanguage}[2]{{%
\expandafter\ifx\csname l@#1\endcsname\relax
\typeout{** WARNING: IEEEtran.bst: No hyphenation pattern has been}%
\typeout{** loaded for the language `#1'. Using the pattern for}%
\typeout{** the default language instead.}%
\else
\language=\csname l@#1\endcsname
\fi
#2}}
\providecommand{\BIBdecl}{\relax}
\BIBdecl

\bibitem{Basser:1994}
P.~J. Basser, J.~Mattiello, and D.~LeBihan, ``{MR} diffusion tensor
  spectroscopy and imaging,'' \emph{Biophys. J.}, vol.~66, no.~1, pp. 259--267,
  Jan. 1994.

\bibitem{Jahanian:2008}
H.~Jahanian, A.~Yazdan-Shahmorad, and H.~Soltanian-Zadeh, ``4{D} wavelet noise
  suppression of {MR} diffusion tensor data,'' in \emph{Proc. IEEE Int. Conf.
  Acoust., Speech, Signal Process., ICASSP'2008}, Las Vegas, Nevada, Mar. 2008,
  pp. 509--512.

\bibitem{cheng:2014}
J.~Cheng, D.~Shen, and P.-T. Yap, ``Designing single- and multiple-shell
  sampling schemes for diffusion {MRI} using spherical code,'' in \emph{Med.
  Image Comput. Comput. Assist. Interv., MICCAI'2014}, Boston, Massachusetts,
  2014, vol. 8675, pp. 281--288.

\bibitem{ye:2012}
W.~Ye, S.~Portnoy, A.~Entezari, S.~J. Blackband, and B.~C. Vemuri, ``An
  efficient interlaced multi-shell sampling scheme for reconstruction of
  diffusion propagators,'' \emph{{IEEE} Trans. Med. Imag.}, vol.~31, no.~5, pp.
  1043--1050, May 2012.

\bibitem{daducci:2011}
A.~Daducci, J.~D. McEwen, D.~V.~D. Ville, J.~P. Thiran, and Y.~Wiaux,
  ``Harmonic analysis of spherical sampling in diffusion {MRI},'' in
  \emph{Proc. 19th Ann. Meet. Int. Soc. Magn. Reson. Med.}, Jun. 2011.

\bibitem{assemlal:2009}
H.~E. Assemlal, D.~Tschumperl\'{e}, and L.~Brun, ``Evaluation of q-space
  sampling strategies for the diffusion magnetic resonance imaging,'' in
  \emph{Med. Image Comput. Comput. Assist. Interv., MICCAI'2009}, London, UK,
  2009, vol.~12, no. Pt 2, pp. 406--414.

\bibitem{tournier:2013}
J.-D. Tournier, F.~Calamante, and A.~Connelly,
  ``\BIBforeignlanguage{en}{Determination of the appropriate b value and number
  of gradient directions for high-angular-resolution diffusion-weighted
  imaging},'' \emph{\BIBforeignlanguage{en}{{NMR} Biomed.}}, vol.~26, no.~12,
  pp. 1775--1786, Dec. 2013.

\bibitem{haldar:2012}
J.~Haldar and R.~Leahy, ``New linear transforms for data on a fourier 2-sphere
  with application to diffusion {MRI},'' in \emph{Proc. 9th {IEEE} Int. Symp.
  Biomed. Imag., ISBI'2012}, Barcelona, Spain, May 2012, pp. 402--405.

\bibitem{descoteaux:2007}
M.~Descoteaux, E.~Angelino, S.~Fitzgibbons, and R.~Deriche, ``Regularized,
  fast, and robust analytical q-ball imaging,'' \emph{Magn. Reson. Med.},
  vol.~58, no.~3, pp. 497--510, Sep. 2007.

\bibitem{anderson:2005}
A.~W. Anderson, ``Measurement of fiber orientation distributions using high
  angular resolution diffusion imaging,'' \emph{Magn. Reson. Med.}, vol.~54,
  no.~5, pp. 1194--1206, Nov. 2005.

\bibitem{Kennedy-book:2013}
R.~A. Kennedy and P.~Sadeghi, \emph{Hilbert Space Methods in Signal
  Processing}.\hskip 1em plus 0.5em minus 0.4em\relax Cambridge, UK: Cambridge
  University Press, Mar. 2013.

\bibitem{hess:2006}
C.~P. Hess, P.~Mukherjee, E.~T. Han, D.~Xu, and D.~B. Vigneron,
  ``\BIBforeignlanguage{eng}{Q-ball reconstruction of multimodal fiber
  orientations using the spherical harmonic basis},''
  \emph{\BIBforeignlanguage{eng}{Magn. Reson. Med.}}, vol.~56, no.~1, pp.
  104--117, Jul. 2006.

\bibitem{jones:1999}
D.~K. Jones, M.~A. Horsfield, and A.~Simmons, ``Optimal strategies for
  measuring diffusion in anisotropic systems by magnetic resonance imaging,''
  \emph{Magn. Reson. Med.}, vol.~42, no.~3, pp. 515--525, Sep. 1999.

\bibitem{tuch:2004}
D.~S. Tuch, ``Q-ball imaging,'' \emph{Magn. Reson. Med.}, vol.~52, no.~6, pp.
  1358--1372, 2004.

\bibitem{fejes-toth:1949}
L.~Fejes-T\'{o}th, ``On the densest packing of spherical caps,'' \emph{Amer.
  Math. Monthly}, vol.~56, pp. 330--331, 1949.

\bibitem{papadakis:2000}
N.~G. Papadakis, C.~D. Murrills, L.~D. Hall, C.~L. Huang, and
  T.~Adrian~Carpenter, ``Minimal gradient encoding for robust estimation of
  diffusion anisotropy,'' \emph{Magn. Reson. Med.}, vol.~18, no.~6, pp.
  671--679, Jul. 2000.

\bibitem{caruyer:2013}
E.~Caruyer, C.~Lenglet, G.~Sapiro, and R.~Deriche, ``Design of multishell
  sampling schemes with uniform coverage in diffusion {MRI},'' \emph{Magn.
  Reson. Med.}, vol.~69, no.~6, pp. 1534--1540, Jun. 2013.

\bibitem{McEwen2011}
J.~D. McEwen and Y.~Wiaux, ``A novel sampling theorem on the sphere,''
  \emph{{IEEE} Trans. Signal Process.}, vol.~59, no.~12, pp. 5876--5887, Dec.
  2011.

\bibitem{caruyer:thesis}
E.~Caruyer, ``{Q}-space diffusion {MRI}: acquisition and signal processing,''
  these, Universit\'{e} Nice Sophia Antipolis, Jul. 2012.

\bibitem{Khalid:2014}
Z.~Khalid, R.~A. Kennedy, and J.~D. McEwen, ``An optimal-dimensionality
  sampling scheme on the sphere with fast spherical harmonic transforms,''
  \emph{{IEEE} Trans. Signal Process.}, vol.~62, no.~17, pp. 4597--4610, Sep.
  2014.

\bibitem{Sakurai:1994}
J.~J. Sakurai, \emph{Modern Quantum Mechanics}, 2nd~ed.\hskip 1em plus 0.5em
  minus 0.4em\relax Reading, MA: Addison Wesley Publishing Company, Inc., 1994.

\bibitem{caruyer:2012}
E.~Caruyer and R.~Deriche, ``Diffusion {MRI} signal reconstruction with
  continuity constraint and optimal regularization,'' \emph{Med. Image Anal.},
  vol.~16, no.~6, pp. 1113--1120, Aug. 2012.

\bibitem{koay:2011}
C.~G. Koay, ``A simple scheme for generating nearly uniform distribution of
  antipodally symmetric points on the unit sphere,'' \emph{J. Comput. Sci.},
  vol.~2, no.~4, pp. 377--381, Dec. 2011.

\end{thebibliography}

\end{document}